\documentclass[a4paper,fleqn,usenatbib]{mnras}

\usepackage{newtxtext,newtxmath}

\usepackage[T1]{fontenc}
\usepackage{ae,aecompl}

\usepackage{graphicx}	
\usepackage{amsmath}	
\usepackage{amssymb}	
\usepackage{color}

\newcommand{\kms}{\mbox{${\rm km\,s}^{-1}$}}
\newcommand{\ms}{\mbox{${\rm m\,s}^{-1}$}}

\newcommand{\Msolar}{\mbox{${M}_{\sun}$}}
\newcommand{\Rsolar}{\mbox{${R}_{\sun}$}}
\newcommand{\Vsini}{\mbox{${\rm V}_{\rm rot}\sin i_{\ast}$}}
\newcommand{\Mast}{\mbox{${M}_{\ast}$}}
\newcommand{\Rast}{\mbox{${R}_{\ast}$}}

\newcommand{\Mjup}{\mbox{${M}_{J}$}}
\newcommand{\rhoast}{\mbox{$\rho_{\ast}$}}

\newcommand{\rhosun}{\mbox{$\rho_{\sun}$}}
\newcommand{\Rjup}{\mbox{${R}_{J}$}}
\newcommand{\rhojup}{\mbox{$\rho_{J}$}}
\newcommand{\teff}{\mbox{$T_{\rm eff}$}}
\newcommand{\logg}{\mbox{$\log g$}}

\newcommand\T{\rule{0pt}{2.2ex}}

\title[WASP-147b, 160Bb, 164b and 165b]{WASP-147b, 160Bb, 164b and 165b: two hot Saturns and two Jupiters, including two planets with metal-rich hosts}

\author[M. Lendl et al.]{M. Lendl$^{1,2}$\thanks{E-mail: monika.lendl@oeaw.ac.at},
D.R. Anderson$^{3}$,
A. Bonfanti$^{4}$,
F. Bouchy$^{2}$,
A. Burdanov$^{4}$,\newauthor
A. Collier Cameron$^{5}$,
L. Delrez$^{6,4}$,
M. Gillon$^{4}$,
C. Hellier$^{3}$,
E. Jehin$^{4}$,\newauthor
P.F.L. Maxted$^{3}$,
L. Dyregaard Nielsen$^{2}$,
F. Pepe$^{2}$,
D. Pollacco$^{7}$,
D. Queloz$^{2,6}$,\newauthor
D. S\'egransan$^{2}$,
J. Southworth$^{3}$,
B. Smalley$^{3}$,
S. Thompson$^{6}$,
O. Turner$^{2}$,\newauthor
A.H.M.J. Triaud$^{8}$,
S. Udry$^{2}$,
and R.G. West$^{7}$
\\
$^{1}$Space Research Institute, Austrian Academy of Sciences, Schmiedlstr. 6, 8042 Graz, Austria\\
$^{2}$Observatoire de Gen\`eve, Universit\'e de Gen\`eve, Chemin des maillettes 51, 1290 Sauverny, Switzerland\\
$^{3}$Astrophysics Group, Keele University, Staffordshire, ST5 5BG, UK\\
$^{4}$Space sciences, Technologies and Astrophysics Research (STAR) Institute, Universit{\'e} de Li{\`e}ge, All{\'e}e du 6 Ao{\^u}t 17, 4000 Li{\`e}ge, Belgium\\
$^{5}$SUPA, School of Physics and Astronomy, University of St.\ Andrews, North Haugh, Fife, KY16 9SS, UK\\
$^{6}$Cavendish Laboratory, J J Thomson Avenue, Cambridge, CB3 0HE, UK\\
$^{7}$Department of Physics, University of Warwick, Gibbet Hill Road, Coventry CV4 7AL, UK\\
$^{8}$School of Physics \& Astronomy, University of Birmingham, Edgbaston, Birmingham B15 2TT, UK\\
}

\date{Accepted XXX. Received YYY; in original form ZZZ}

\pubyear{2018}

\begin{document}
\label{firstpage}
\pagerange{\pageref{firstpage}--\pageref{lastpage}}
\maketitle

\begin{abstract}
We report the discovery of four transiting hot Jupiters, WASP-147, WASP-160B, WASP-164 and WASP-165 from the WASP survey. WASP-147b is a near Saturn-mass 
($M_P = 0.28 \Mjup$) object with a radius of $ 1.11 \, \Rjup$ orbiting a G4 star with a period of $ 4.6 $~d. 
WASP-160Bb has a mass and radius ($ M_p = 0.28 \, \Mjup $, $ R_p = 1.09 \, \Rjup$) near-identical to WASP-147b, but is less irradiated, 
orbiting a metal-rich ($\mathrm{[Fe/H]}_{\ast}=0.27$) K0 star with a period of $3.8$~d. WASP-160B is part of a near equal-mass visual binary with an on-sky separation of 28.5~arcsec. 
WASP-164b is a more massive ($M_P = 2.13 \, \Mjup$, $R_p = 1.13 \, \Rjup$) hot Jupiter, orbiting a G2 star on a close-in ($P=1.8$~d), 
but tidally stable orbit. WASP-165b is a classical ($M_p = 0.66 \, \Mjup$, $R_P = 1.26 \, \Rjup$) hot Jupiter in a $3.5$~d 
period orbit around a metal-rich ($\mathrm{[Fe/H]_{\ast}}=0.33$) star. 
WASP-147b and WASP-160Bb are promising targets for atmospheric characterization through transmission spectroscopy, while WASP-164b 
presents a good target for emission spectroscopy.
\end{abstract}




\section{Introduction}


Transiting exoplanets are invaluable objects for study. Not only are both, their masses and radii known, but also their transiting configuration 
opens up a wide range of characterization avenues. We may study the atmospheres of these objects through their transmission and emission spectra 
\citep{Seager00,Charbonneau02,Charbonneau05}, but also measure their orbital alignment \citep{Queloz00}; see \citet{Triaud17} 
for a summary. While over 2700 transiting planets are known to date\footnote{according to \texttt{exoplanets.eu}, queried on 12 Apr 2018.}, only 
a fraction of these objects are suitable for detailed characterization, as this requires the planet host to be bright and the star/planet size ratio to be favorable.

Ground based transit surveys (e.g. WASP, \citealp{Pollacco06}; HAT, \citealp{Bakos04}; KELT, \citealp{Pepper07}; and MASCARA,\citealp{Talens17}) 
use small-aperture instrumentation to monitor vast numbers of bright stars across nearly the entire sky, sensitive to the $\sim$~1\% dips created by transits of 
close-in giant planets. These \emph{hot Jupiters} are prime targets for further characterization thanks to their large radii, frequent transits and extended atmospheres. 
Indeed, ground-based transit surveys have provided some of the most intensely studied planets to date (e.g.\ WASP-12b, \citealt{Hebb09}; WASP-43b, \citealt{Hellier11b} and HAT-P-11b, 
\citealt{Bakos10}). 

In this paper, we report the discovery of four additional close-in transiting gas giants by WASP-South, the two Saturn-mass planets WASP-147b and WASP-160Bb, and the 
two hot Jupiters WASP-164b and WASP-165b. We discuss the observations leading to these discoveries in Section \ref{sec:obs}, describe their host stars in Section \ref{sec:stel} 
and discuss the individual planetary systems and their place among the known planet population in Section \ref{sec:sys} before concluding in Section \ref{sec:sum}.

\section{Observations}
\label{sec:obs}

WASP-147 (2MASS 23564597-2209113), WASP-160B (2MASS~05504305-2737233), WASP-164 (2MASS~22592962-6026519) and WASP-165 (2MASS~23501932-1704392) were monitored with the 
WASP-South facility throughout several years between 2006 and 2013. In the case of WASP-160B, the target flux was blended with that of another object (2MASS~05504470-2737050, 
revealing to be physically associated, see below) in the WASP aperture. The WASP-South instrument consists of an array of 8 cameras equipped with 200mm f/8 Canon lenses on a 
single mount and is located at SAAO (South Africa). For details on observing strategy, data reduction and target selection, please refer to \citet{Pollacco06} and \citet{Cameron07}. 
Using the algorithms described by \citet{Cameron06}, we identified periodic flux drops compatible with transits of close-in giant planets in the light curves of these four objects. 
We thus triggered spectroscopic and photometric follow-up observations to determine the nature of the observed dimmings.

\subsection{Follow-up spectroscopy}
\begin{figure}
	\includegraphics[width=\columnwidth]{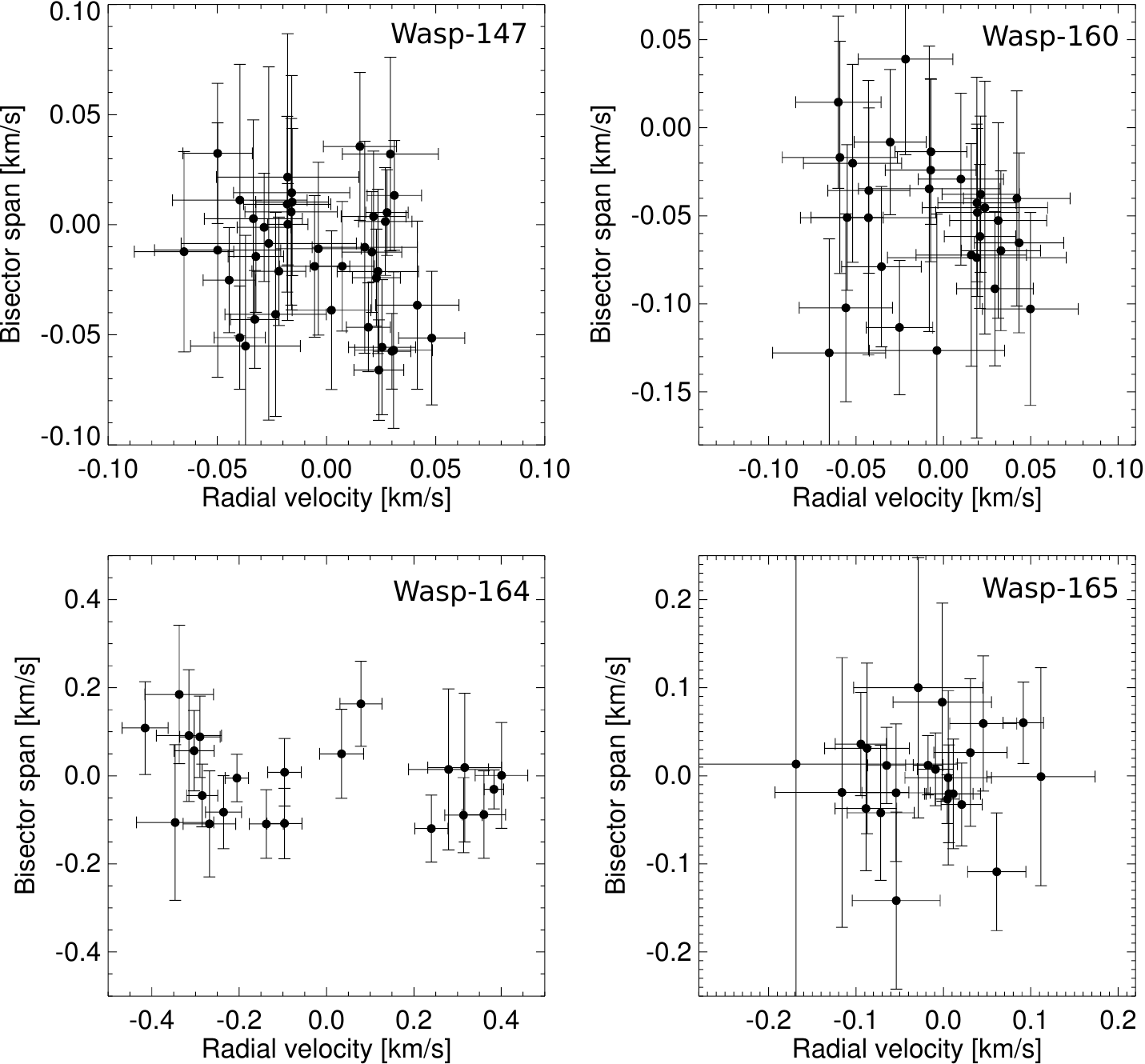}
    \caption{\label{fig:bis}Bisector spans against RV of our targets. The RVs have been corrected for the systemic velocities given in Table \ref{tab:par}.}
    \label{fig:example_figure}
\end{figure}

\begin{figure}
	\includegraphics[width=\columnwidth]{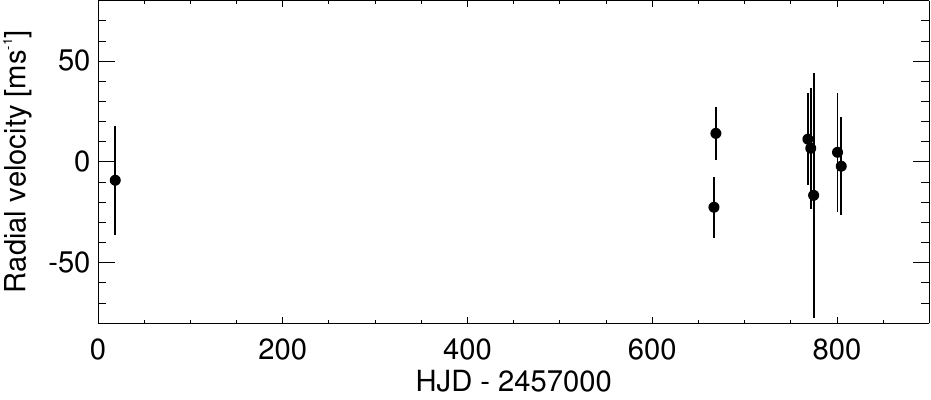}
    \caption{\label{fig:W160A_RV}CORALIE RVs of WASP-160A.}
\end{figure}

We obtained spectroscopic observations of all four objects using the CORALIE echelle spectrograph at the 1.2m Euler-Swiss telescope at La~Silla. From the spectra, we computed 
radial velocities (RVs) using the weighted cross-correlation method \citep{Baranne96,Pepe02}. In 2014, the CORALIE spectrograph was upgraded by replacing circular with octagonal 
fibers, leading to a shift in RV zero point between observations obtained before and after the exchange. WASP-160B, 164 and 165 were observed only after the upgrade, resulting 
in a single homogeneous set of RVs for each object. WASP-147 was observed before and after the upgrade, making it necessary to include these observations as two separate data sets in our analysis. 
For each of the four objects in question, RV variations confirmed the presence of a planet orbiting at the period of the observed transits (see Figs. \ref{fig:W147} -- \ref{fig:W165}).
The individual RV measurements are listed in Tables \ref{tab:RV147} -- \ref{tab:RV165}. 
To exclude stellar activity as the origin of the observed RV variations, we verified that bisector spans and RVs are uncorrelated (\citealp{Queloz01}; 
Pearson coefficients are $-0.19\pm0.16$, $-0.16\pm0.19$, $-0.24\pm0.22$ and $0.06\pm0.22$ for WASP-147, 160B, 164 and 165, respectively). This is illustrated in Fig. \ref{fig:bis}, where we plot bisector spans against RVs. 
As both stellar components of the WASP-160AB system fell into the same WASP-South aperture, we could not a-priori exclude either of them as the origin of the observed transits. 
We thus obtained several spectra of WASP-160A, showing no evidence of any large-amplitude RV variability (see Fig. \ref{fig:W160A_RV}).

\begin{table*}
	\centering
	\caption{\label{tab:RV147} RV measurements of WASP-147. Only the first five entries are shown; the full table is available in the online version.
	}
	\begin{tabular}{ccccc} 
		\hline
		HJD-2450000 & RV [{\kms}] & error [{\kms}] & bisector [{\kms}] & note \\
		\hline
		6135.904214 & -1.70365 & 0.02277 & -0.01228 & pre-upgrade \\
		6137.920709 & -1.62092 & 0.02408 & -0.01027 & pre-upgrade \\
		6157.712465 & -1.67809 & 0.03085 & 0.01114  & pre-upgrade \\
		6158.661748 & -1.65414 & 0.01670 & 0.01027  & pre-upgrade \\
		6508.712267 & -1.65458 & 0.02119 & 0.00580  & pre-upgrade \\
		\hline
	\end{tabular}
\end{table*}

\begin{table}
	\centering
	\caption{\label{tab:RV160B} RV measurements of WASP-160B. Only the first five entries are shown; the full table is available in the online version.
	}
	\begin{tabular}{cccc} 
		\hline
		HJD-2450000 & RV [{\kms}] & error [{\kms}] & bisector [{\kms}] \\
		\hline
		7011.653767 & -6.16374 & 0.02709 &  0.03897 \\
		7016.726245 & -6.12609 & 0.03159 & -0.07226 \\
		7019.639217 & -6.15020 & 0.04051 & -0.03465 \\
		7037.692732 & -6.20130 & 0.03296 & -0.01686 \\
		7038.718708 & -6.11831 & 0.03588 & -0.04539 \\
		\hline
	\end{tabular}
\end{table}

\begin{table}
	\centering
	\caption{\label{tab:RV164} RV measurements of WASP-164. Only the first five entries are shown; the full table is available in the online version.
	}
	\begin{tabular}{cccc} 
		\hline
		HJD-2450000 & RV [{\kms}] & error [{\kms}] & bisector [{\kms}] \\
		\hline
		7185.833271 & 12.66344 & 0.06016 & 0.00090  \\
		7191.863118 & 11.92547 & 0.07861 & 0.18480  \\
		7192.836522 & 12.62296 & 0.04957 & -0.08816 \\
		7193.822033 & 11.95951 & 0.04570 & 0.05691  \\
		7200.871113 & 11.97277 & 0.04621 & 0.08865  \\
		\hline
	\end{tabular}
\end{table}

\begin{table}
	\centering
	\caption{\label{tab:RV165} RV measurements of WASP-165. Only the first five entries are shown; the full table is available in the online version.
	}
	\begin{tabular}{cccc} 
		\hline
		HJD-2450000 & RV [{\kms}] & error [{\kms}] & bisector [{\kms}] \\
		\hline
		7014.541384 & 25.74712 & 0.02309 & 0.06029  \\
		7016.541690 & 25.56122 & 0.02932 & 0.03612  \\
		7178.889363 & 25.65436 & 0.05627 & 0.08364  \\
		7187.897899 & 25.70129 & 0.03835 & 0.05943  \\
		7190.895590 & 25.66115 & 0.04947 & -0.00215 \\
		\hline
	\end{tabular}
\end{table}

\subsection{Follow-up photometry}
\begin{figure}
	\includegraphics[width=\columnwidth]{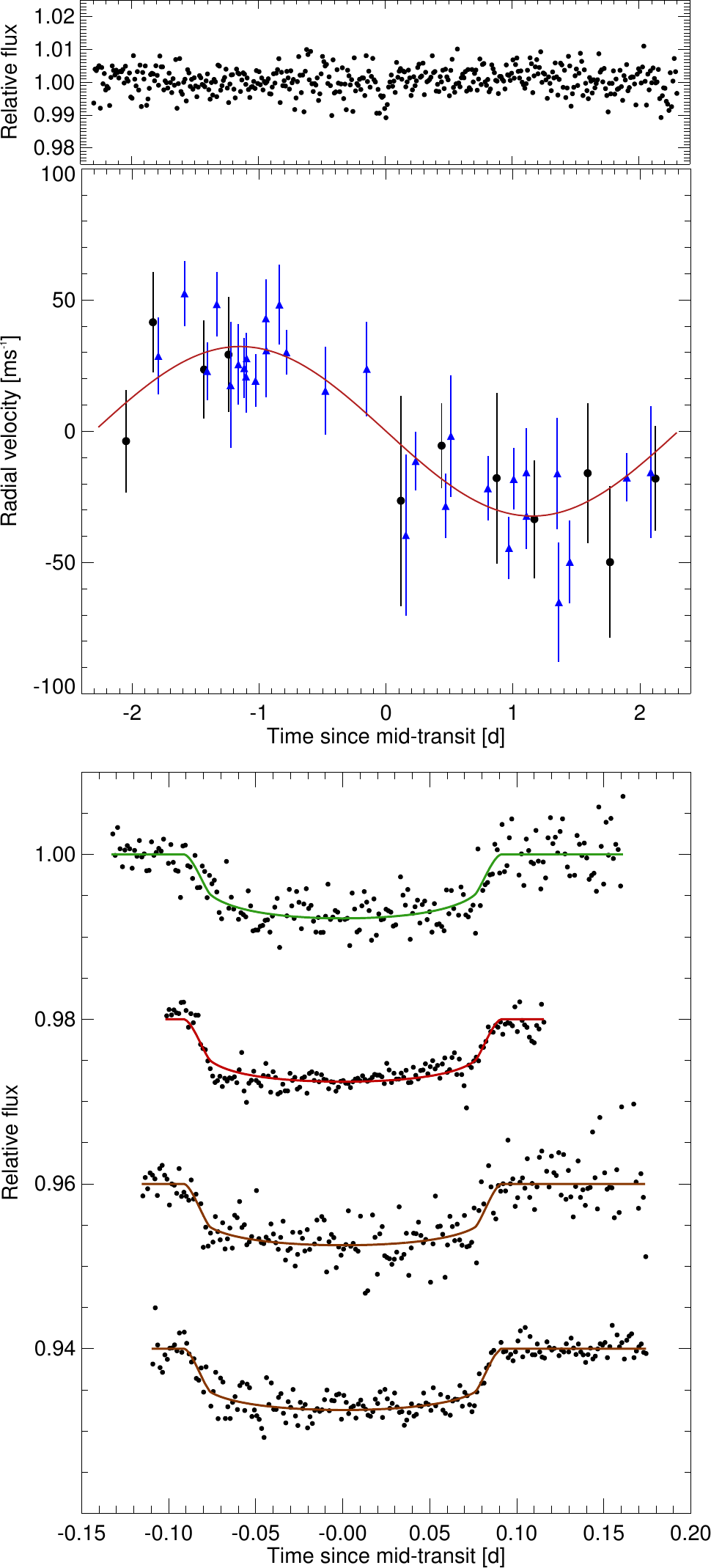}
    \caption{\label{fig:W147}Discovery and follow-up photometry and RVs of WASP-147. Top panel: WASP survey data, 
    phase-folded on the period of WASP-147b and binned per 15 minutes. Middle panel: CORALIE RV data, where the 
    pre-upgrade data are shown as blue triangles, and the post-upgrade data are shown as black filled circles. Bottom panel: 
    Follow-up transit light curves, corrected for their respective baseline models and binned by two minutes. 
    They are (from top to bottom):
    V-band TRAPPIST data of 05 Nov 2014, and r'-band EulerCam data of 11 Nov 2013, and blue-block TRAPPIST data of
    19 Oct 2013 and 26 Sep 2013. The systematics seen in the 11 Nov 2013 light curve are due to cloud passages.}
\end{figure}

\begin{figure}
	\includegraphics[width=\columnwidth]{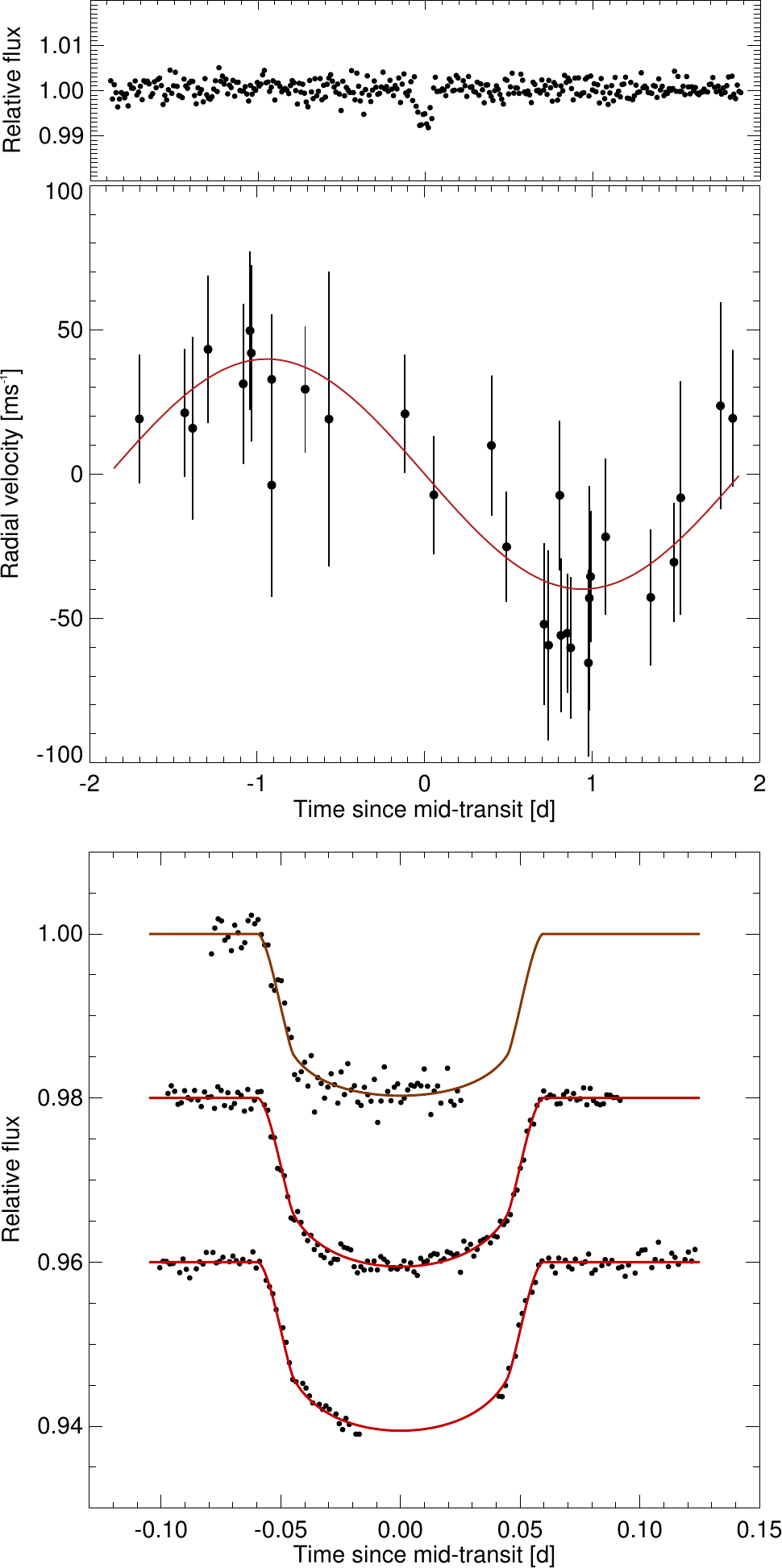}
    \caption{\label{fig:W160}Discovery and follow-up photometry and RVs of WASP-160B. As Fig.\ \ref{fig:W147}.
    The light curves shown are (from top to bottom): I+z'-band TRAPPIST data of 16 Dec 2014, and 
    r'-band EulerCam data of 26 Dec 2015 and 02 Jan 2017. Note that the transit depth in the WASP light curve is 
    reduced due to contamination from WASP-160A.}
\end{figure}

\begin{figure}
	\includegraphics[width=\columnwidth]{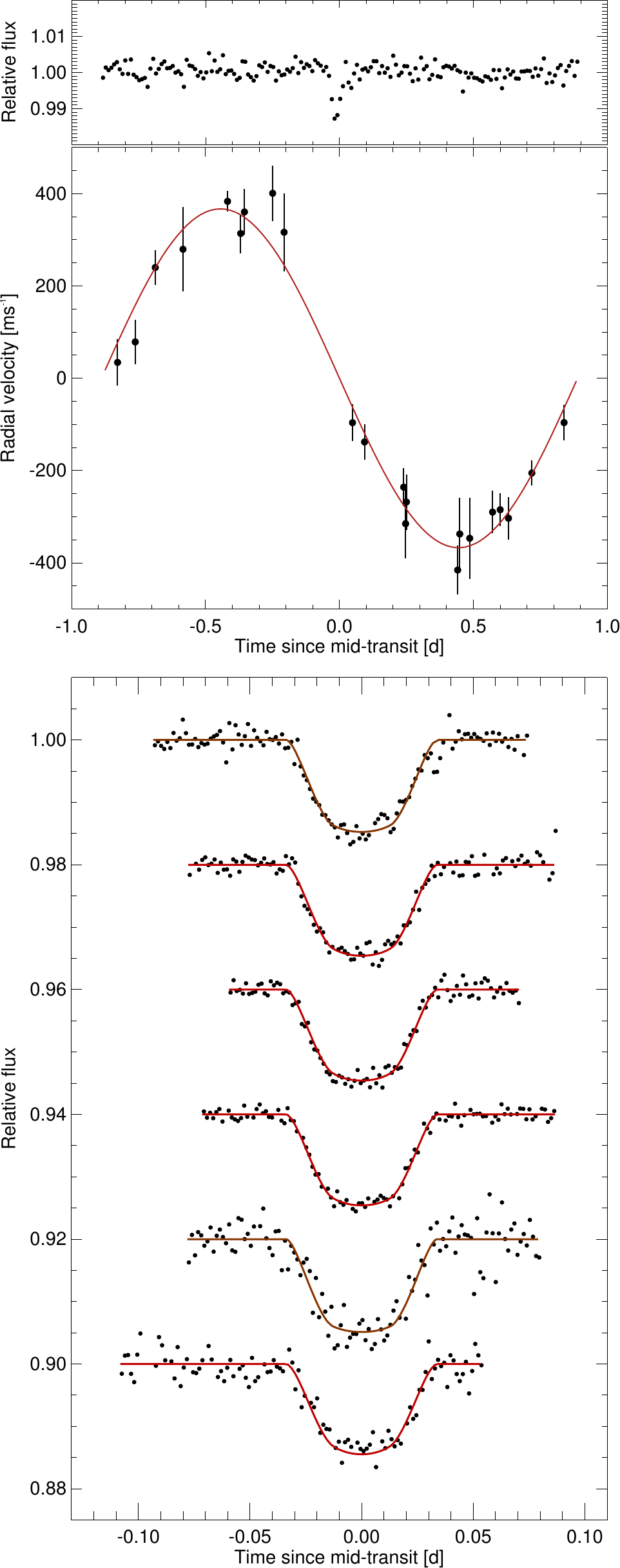}
    \caption{\label{fig:W164}Discovery and follow-up photometry and RVs of WASP-164. As Fig.\ \ref{fig:W147}.
    The light curves shown are (from top to bottom): blue-block TRAPPIST data of 29 Jun 2015, r'-band 
    EulerCam data of 31 Jul 2015, 16 Aug 2015 and 25 Aug 2015, I+z'-band TRAPPIST data of 10 Sep 2016 and
    and R-band SAAO data of 16 Oct 2016.}
\end{figure}

\begin{figure}
	\includegraphics[width=\columnwidth]{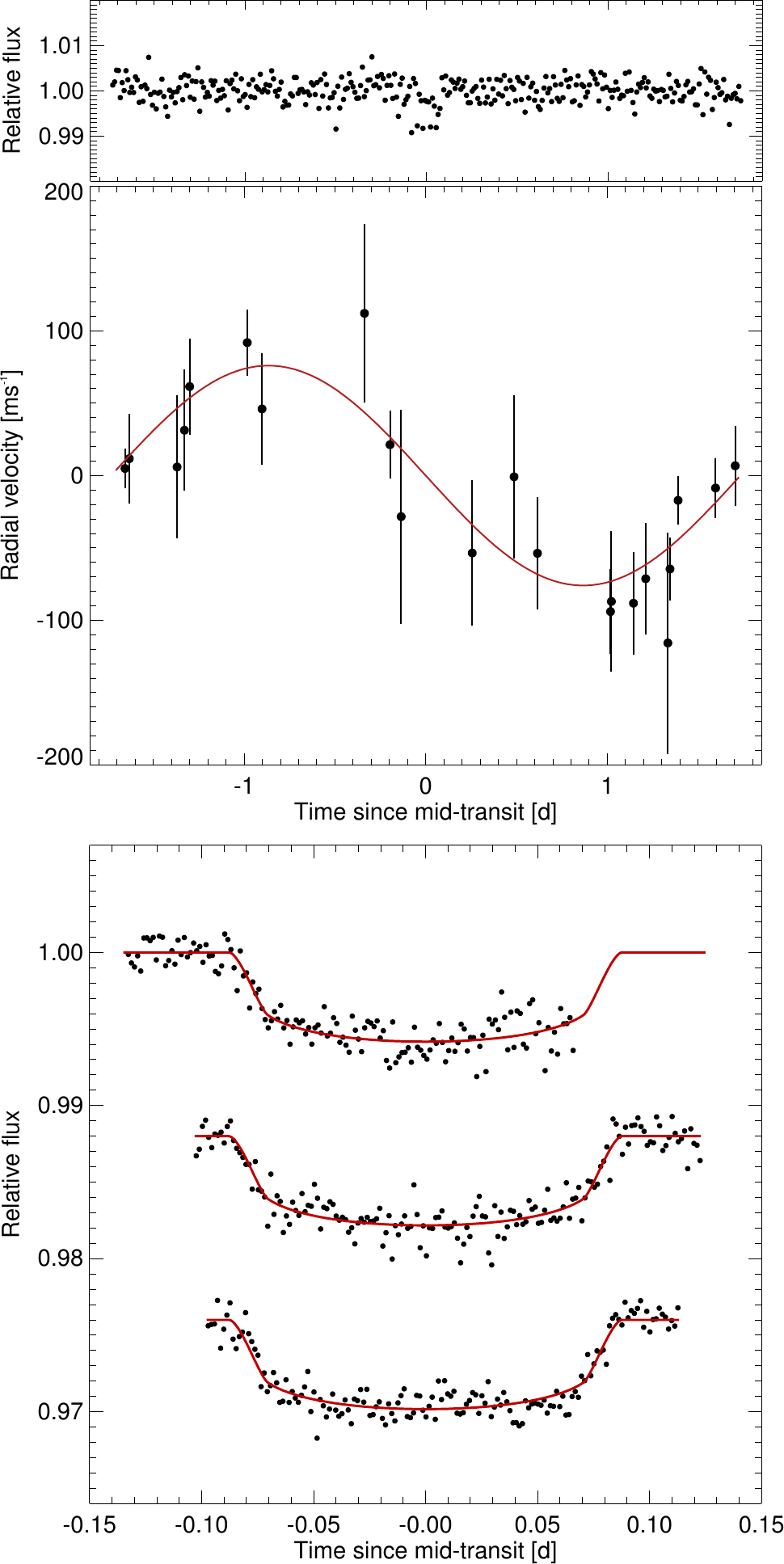}
    \caption{\label{fig:W165}Discovery and follow-up photometry and RVs of WASP-165. As Fig.\ \ref{fig:W147}.
    The light curves shown are (from top to bottom): NGTS-filter EulerCam light curves of 10 Mar 2015, 17 Sep 2016 and 24 Sep 2016.}
\end{figure}

We obtained several high-precision transit light curves for each of our targets to obtain an improved measurement of the transit shape and depth. The facilities we used for 
this purpose were EulerCam at the 1.2m Euler-Swiss telescope \citep{Lendl12}, the 0.6m TRAPPIST-South telescope 
\citep{Gillon11a, Jehin11}, and the SAAO 1.0m telescope. In all cases, we extracted light curves of the transit events using relative aperture photometry,
while iteratively selecting reference stars and aperture sizes to minimize the final light curve RMS. Having an on-sky separation of 
$28.478948 \pm 2.5\times10^{-5}$~arcsec, both stellar components of the WASP-160 system were well-separated in these observations, confirming the fainter star, 
WASP-160B as the origin of the transit feature. Details on all photometric follow-up observations are listed in Table \ref{tab:obs}. The resulting light curves are shown in 
Figs.\ \ref{fig:W147}--\ref{fig:W165}.

\begin{table*}
	\centering
	\caption{\label{tab:obs}Summary of photometric follow-up observations together with the preferred baseline model, noise correction factors and the light curves' RMS per 5 minute bin. 
	The notation of the baseline models, \textit{p(j\textsuperscript{i})}, refers to a polynomial of degree i in parameter j. Filter ranges:
$\lambda_{\mathit{blue-block}} > 500$~nm, $\lambda_{\mathit{NGTS}} = [500 - 900]$~nm, \citet{Wheatley18}. 
	}
	\begin{tabular}{ccccccc} 
		\hline
		Date & Instrument & Filter & Baseline & $\beta_{r}$ & $\beta_{w}$ & $\mathrm{RMS_{5min}}$ [ppm] \\
		\hline
		\multicolumn{7}{l}{WASP-147} \\
		\hline
		05 Nov 2014 & TRAPPIST-South & V & $p(t^{2})$ & 1.22 & 1.22 & 1419 \\
		11 Nov 2013 & EulerCam & r'& $p(t^{2})+p(FWHM^{1})$ &  2.42 & 1.64 & 1020 \\
		19 Oct 2013 & TRAPPIST-South & \emph{blue-block} & $p(t^{2})$ &  1.33 & 0.98 & 1682 \\
		26 Sep 2013 & TRAPPIST-South & \emph{blue-block} & $p(t^{2})$ &  2.02 & 0.81 & 900 \\
		\hline
		\multicolumn{7}{l}{WASP-160B} \\
		\hline
		16 Dec 2014 & TRAPPIST-South & I+z' & $p(t^{2})$ & 1.16 & 0.99 & 1091 \\
		26 Dec 2015 & EulerCam & r' & $p(t^{2})$ & 1.50 & 1.06 & 469 \\
		02 Jan 2017 & EulerCam & r' & $p(t^{2})$ & 1.92 & 1.08 & 628 \\
		\hline
		\multicolumn{7}{l}{WASP-164} \\
		\hline
		29 Jun 2015 & TRAPPIST-South & \emph{blue-block} & $p(t^{2})$  &  1.58 & 0.89 & 1070 \\
		31 Jul 2015 & EulerCam & r' & $p(t^{2})+p(sky^{1})$ &  1.21 & 1.52 & 1103 \\
		16 Aug 2015 & EulerCam & r' & $p(t^{2})$  &  1.00 & 1.45 & 616 \\
		25 Aug 2015 & EulerCam & r' & $p(t^{2})+p(xy^{1})$ &  1.09 & 1.31 & 503 \\
		10 Sep 2016 & TRAPPIST-South & I+z' & $p(t^{2})+p(FWHM^{2})+p(xy^{1})$  &  1.01 & 2.81 & 1301 \\
		16 Oct 2016 & SAAO 1m & R & $p(t^{2})$  &  1.61 & 0.85 & 1302 \\
		\hline
		\multicolumn{7}{l}{WASP-165} \\
		\hline
		10 Mar 2015 & EulerCam & \emph{NGTS} & $p(t^{2})+p(FWHM^{1})$ &  1.52 & 1.60 & 623 \\
		17 Sep 2016 & EulerCam & \emph{NGTS} & $p(t^{2})+p(FWHM^{1})$ &  1.32 & 1.46 & 626 \\
		24 Sep 2016 & EulerCam & \emph{NGTS} & $p(t^{2})+p(FWHM^{1})$ &  1.98 & 1.41 & 559 \\		
		\hline
	\end{tabular}
\end{table*}

\section{Stellar parameters}
\label{sec:stel}

\subsection{Spectral analysis}
\label{sec:specana}

The individual CORALIE spectra for each star were co-added in order to provide a spectrum for analysis. Using methods similar to those described by \cite{2013MNRAS.428.3164D}, 
for each star we determined the effective temperature ($T_\mathrm{eff}$), surface gravity ($\log g$), stellar metallicity ([Fe/H]), and projected stellar rotational velocity 
($v_{\rm rot} \sin i_\ast$). In determining $v_{\rm rot} \sin i_\ast$ we assumed a macroturbulent velocity using the calibration given by \citep{2014MNRAS.444.3592D}. 
For WASP-160B and WASP-165 the $v_{\rm rot} \sin i_\ast$ values are consistent with zero and that of WASP-147 can be considered an upper limit, as it is close to the resolution limit of the CORALIE spectrograph. If, however, a zero macroturbulent 
velocity is used, we obtain $v_{\rm rot} \sin i_\ast$ values of 3.1~$\pm$~0.5, 0.7~$\pm$~0.6 and 2.8~$\pm$~0.6~km\,s$^{-1}$ for WASP-147, WASP-160B and WASP-165, respectively.

The parameters for WASP-164 are relatively poorly determined, as the signal-to-noise ratio of the merged spectrum for WASP-164 is very low (\la 20:1). 
The $v_{\rm rot} \sin i_\ast$ is consistent with zero, but very poorly determined and the Lithium 670.8\,nm line might be present, but we cannot be sure due to the quality of the spectrum.

\begin{table*}
\centering                        
\caption{\label{tab:stel}Basic properties and stellar parameters of the planet hosts based on spectroscopic analysis, evolutionary models and photometric variability. 
$^{a}$Derived from Gaia DR2 parallaxes \citep{gaia18} corrected for the -80 $\mu$arcsec offset found by \citet{Stassun18};
$^{b}$Using \citet{Sestito05}; $^{c}$Internal model grid uncertainty; $^{d}$Estimated uncertainty accounting for different model grids;
$^{e}$Using \citet{Barnes07}.}
\begin{tabular}{lcccc}       
\hline\hline 
 Parameter & WASP-147 & WASP-160B & WASP-164 & WASP-165 \T  \\
\hline
\multicolumn{5}{l}{Basic parameters}\\
\hline
RA (J2000)    &  23 56 45.97 &  05 50 43.06 &  22 59 29.62 &  23 50 19.33 \\
DEC (J2000)   & -22 09 11.39 & -27 37 23.39 & -60 26 51.97 & -17 04 39.26 \\
UCAC4 B [mag]       & $12.96\pm0.04 $ & $13.98 \pm0.03 $ & $13.32 \pm 0.03$ & $13.41 \pm0.03 $\\
UCAC4 V [mag]       & $12.31 (\pm <0.01) $ & $13.09 \pm0.01 $ & $12.62 \pm 0.01$ & $12.69 \pm0.04 $\\
2MASS J [mag]       & $11.174\pm0.023$ & $11.591\pm0.030$ & $11.365\pm0.024$ & $11.439\pm0.023$\\
2MASS H [mag]       & $10.907\pm0.025$ & $11.172\pm0.024$ & $11.040\pm0.024$ & $11.125\pm0.024$\\
2MASS K [mag]       & $10.857\pm0.025$ & $11.055\pm0.019$ & $10.959\pm0.021$ & $11.024\pm0.023$\\
GAIA DR2 ID   & 2340919358581488768 & 2910755484609597312 & 6491038642006989056 & 2415410962124813056 \\
GAIA G [mag]       & 12.2 & 12.9 & 12.5 & 12.5 \\
GAIA $G_{\rm BP}$ [mag] & 12.5 & 13.3 & 12.8 & 12.9 \\
GAIA $G_{\rm RP}$ [mag] & 11.7 & 12.3 & 11.9 & 12.0 \\
Distance [pc]$^{a}$ & $426\pm14$ & $284\pm5$ & $322\pm7$& $583\pm34$ \\ 
\hline
\multicolumn{5}{l}{Parameters from spectral analysis} \\
\hline
Spectral type & G4 & K0V & G2V & G6 \\
\teff [K]        & $ 5700 \pm 100 $     & $5300 \pm 100 $   & $5800 \pm 200$ & $5600\pm150$ \\
$\mathrm{[Fe/H]}$      & $ +0.09 \pm 0.07$    & $+0.27 \pm 0.1 $ & $0.0 \pm 0.2 $ & $+0.33\pm0.13 $\\
\logg [cgs]   & $ 4.0 \pm 0.1 $      & $4.6 \pm 0.1 $    & $4.5 \pm 0.2 $ & $3.9\pm0.2$ \\
\Vsini [\kms] & $ 0.3_{-0.3}^{+0.8}$ & $\sim 0$ & $\sim 0$ & $\sim 0$ \\
$\log{A({\rm Li})}$ & $ 1.91 \pm 0.09 $    & - & - & $2.21 \pm 0.14$ \\
Lithium Age$^b$ [Gyr] & $\ge2$ & $\ge 0.5$ & - & $0.5 - 2$ \\
\hline
\multicolumn{5}{l}{Parameters from stellar evolutionary models} \\
\hline
\Mast & $1.04 (\pm0.02)^c \pm 0.07^{d} $ & $0.87 (\pm0.02)^c \pm 0.07^{d} $ & $0.95 (\pm0.04)^c \pm 0.07^{d} $ & $1.25 (\pm0.04)^c \pm 0.07^{d}$ \\
\Rast & $1.37 (\pm0.03)^c \pm 0.07^{d} $ & $0.87 (\pm0.01)^c \pm 0.07^{d} $ & $0.92 (\pm0.02)^c \pm 0.07^{d} $ & $1.65 (\pm0.06)^c \pm 0.07^{d}$ \\
Age [Gyr]  & $ 8.47 \pm 0.78 $ & $ 9.75 \pm 2.28 $ & $4.08 \pm 2.38$ & $4.77 \pm 0.92$ \\
\hline
Gyrochronological age$^e$ [Gyr] & - & - & $2.32^{+0.98}_{-0.55}$ & - \\
\hline
\end{tabular}
\end{table*}

\subsection{Rotation periods}
\begin{figure}
\mbox{\includegraphics[width=0.49\textwidth]{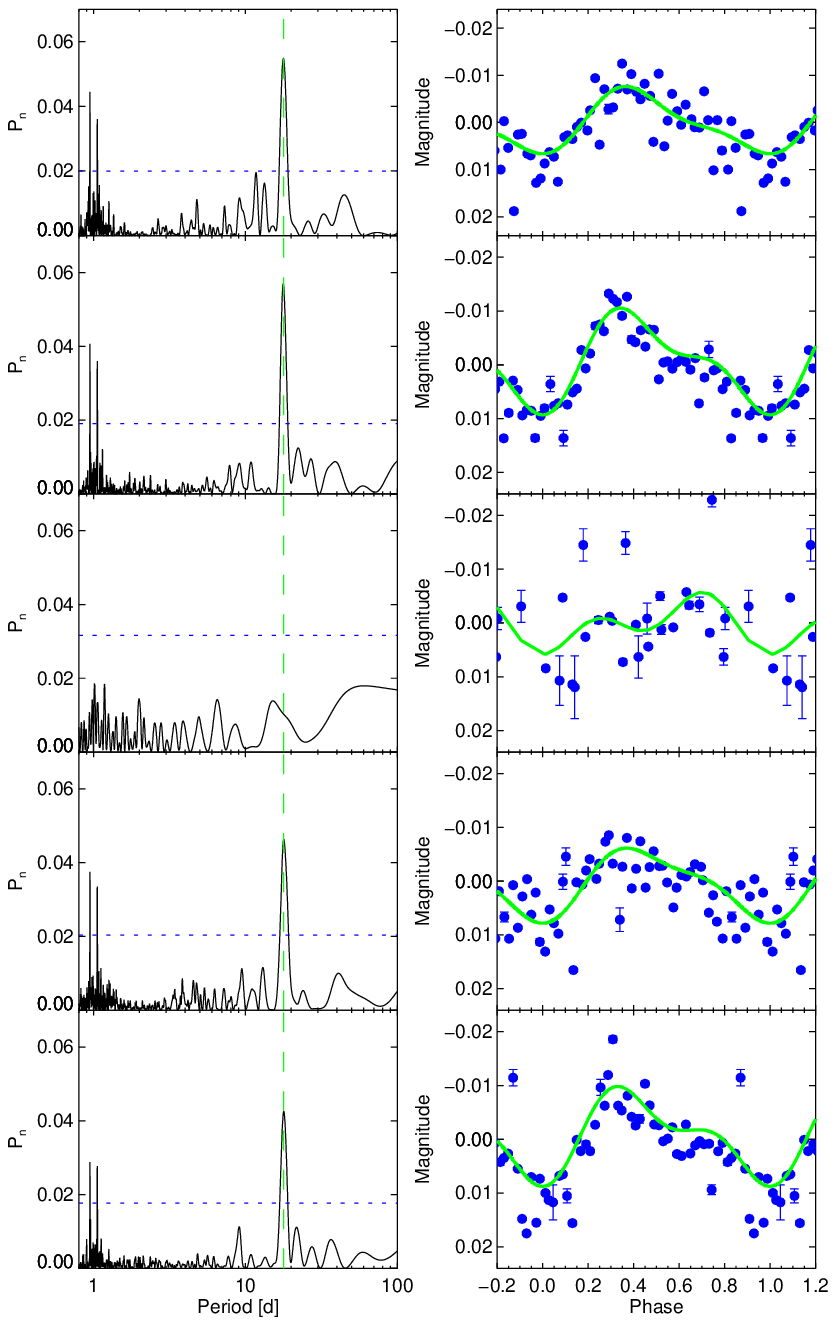}}
\caption{Left: Periodograms of the WASP light curves for WASP-164. Horizontal
lines indicate false-alarm probability levels 0.1, 0.01 and 0.001. Right:
Light curves phase-binned on the assumed rotation period of 17.81\,days (points) with second-order harmonic series fit by least squares (lines). Data are
plotted by season and camera in the same order top-to-bottom as in
Table~\protect\ref{ProtTable}. \label{ProtFig}}
\end{figure}

 The WASP light curves of WASP-164 show a quasi-periodic modulation with an
amplitude of about 0.6 per cent and a period of about 18 days. We assume this
is due to the combination of the star's rotation and magnetic activity, i.e.,
star spots. We used the sine-wave fitting method described in
\citet{Maxted11} to refine this estimate of the amplitude and period
of the modulation. Variability due to star spots is not expected to be coherent
on long timescales as a consequence of the finite lifetime of star-spots and
differential rotation in the photosphere so we analyzed each season of data for
WASP-164 separately. We also analyze the data from each camera used to observe
WASP-164 separately so that we can assess the reliability of the results. We
removed the transit signal from the data prior to calculating the periodograms
by subtracting a simple transit model from the light curve. We calculated
periodograms over 8192 uniformly spaced frequencies from 0 to 1.5 cycles/day.
The false alarm probability (FAP) is calculated using a boot-strap Monte Carlo
method also described in \citet{Maxted11}. The results are given in
Table~\ref{ProtTable} and the periodograms and light curves for are shown in
Fig.~\ref{ProtFig}. There is a clear signal near 17.8 days in 4 out of 5 data
sets, from which we obtain a value for the rotation period of $P_{\rm rot} =
17.81 \pm 0.03$\,d.  This rotation period together with our estimate the stellar
radius implies a value of $V_{\rm rot}\sin I = 2.6 \pm 0.2$\,\kms, assuming
that the rotation axis of the star is approximately aligned with the orbital
axis of the planet. This is consistent with the low
value for $v_{\rm rot}\sin i$ we obtain from our analysis of the spectroscopy
of WASP-164. We used a least-squares fit of a sinusoidal function and its first
harmonic to model the rotational modulation in the light curves for each camera
and season with the rotation period fixed at $P_{\rm rot}  = 17.81$\,d. 

For WASP-147, WASP-160B and WASP-165, a similar analysis leads to upper limits of
1.2\,millimagnitudes, 2.9\, millimagnitudes and 1.2\,millimagnitudes with
95\,per~cent confidence for the amplitude of any sinusoidal signal over the
same frequency range, respectively.

\begin{table}
 \caption{Periodogram analysis of the WASP light curves for WASP-164. Observing
dates are JD-2450000,  $N$ is the number of observations used in the analysis,
$a$ is the semi-amplitude of the best-fit sine wave at the period $P$ found
in the periodogram with false-alarm probability FAP.
\label{ProtTable}}
 \begin{tabular}{@{}rrrrrr}
\hline
  \multicolumn{1}{@{}l}{Camera} &
  \multicolumn{1}{l}{Dates} &
  \multicolumn{1}{l}{$N$} &
  \multicolumn{1}{l}{$P$ [d]} &
  \multicolumn{1}{l}{$a$ [mmag]} &
  \multicolumn{1}{l}{FAP}\\
 \noalign{\smallskip}
\hline
\hline
221 & 5336-5515 &  7916  &  17.790  &  0.006  & $<10^{-4}$ \\
221 & 5699-5881 &  5623  &  17.730  &  0.007  & $<10^{-4}$ \\
221 & 6064-6106 &  1004  &   1.014  &  0.004  & 0.93   \\
222 & 5352-5527 &  7906  &  17.850  &  0.006  & $<10^{-4}$ \\
222 & 5716-5897 &  5041  &  17.850  &  0.007  & $<10^{-4}$ \\
 \noalign{\smallskip}
\hline
 \end{tabular}   
 \end{table}     

\subsection{Stellar evolution modeling}
\label{sec:evol}

In order to estimate stellar parameters, we considered $T_{\mathrm{eff}}$, [Fe/H], $\log{g}$ and $v\sin{i}$ inferred from spectral analysis (see Sec. \ref{sec:specana}), 
the mean stellar density $\rho_{\star}$ inferred from the transit light curve (see Sec. \ref{sec:mod}) and magnitude $G$, color index $G_{BP}-G_{RP}$ and distance $d$ reported 
by Gaia DR2 \citep{gaia18}. We analyzed WASP-147 and WASP-160B using the set $G$, $G_{BP}-G_{RP}$, $d$, [Fe/H], $\log{g}$, $\rho_{\star}$ and $v\sin{i}$ as input parameters. 
For WASP-164 and WASP-165, we adopted the same input set, but replaced the color index by the spectroscopic temperature as it is more precisely known than that inferred from the Gaia colors. 
An extinction coefficient of $Av=0$ was assumed in this analysis.
We recovered the main stellar parameters such as age, mass and radius according to theoretical models. We considered the grids of evolutionary tracks and isochrones computed by PARSEC\footnote{%
Padova and Trieste Stellar Evolutionary Code. http://stev.oapd.inaf.it/cgi-bin/cmd}%
(version 1.2S; see \citet{bressan12,chen14} and references therein).

The interpolation of the input data in the theoretical grids to retrieve the output parameters has been done according to the isochrone placement technique described in 
\citet{bonfanti15,bonfanti16}. Here we briefly recall that the algorithm makes a comparison between observations and theoretical isochrones and select those theoretical 
data which match the observations best. In particular, for each star:
\begin{itemize}
 \item multiple grids of isochrones spanning the input metallicity range [[Fe/H]$-\Delta$[Fe/H]; [Fe/H]$+\Delta$[Fe/H]] have been loaded;
 \item isochrones have been filtered through a 2-dimensional Gaussian window function whose $\sigma_1=\Delta T_{\mathrm{eff}}$, $\sigma_2=\Delta\log{L}$;
 \item isochrones have been weighted evaluating the stellar evolutionary speed in the HR diagram and considering the similarity between theoretical and observational parameters;
 \item the gyrochronological relation by \citet{barnes10} has been used to set a conservative age lower limit to discard unlikely very young isochrones;
 \item element diffusion has been taken into account.
\end{itemize}

\begin{figure*}
	\includegraphics[width=0.9\linewidth]{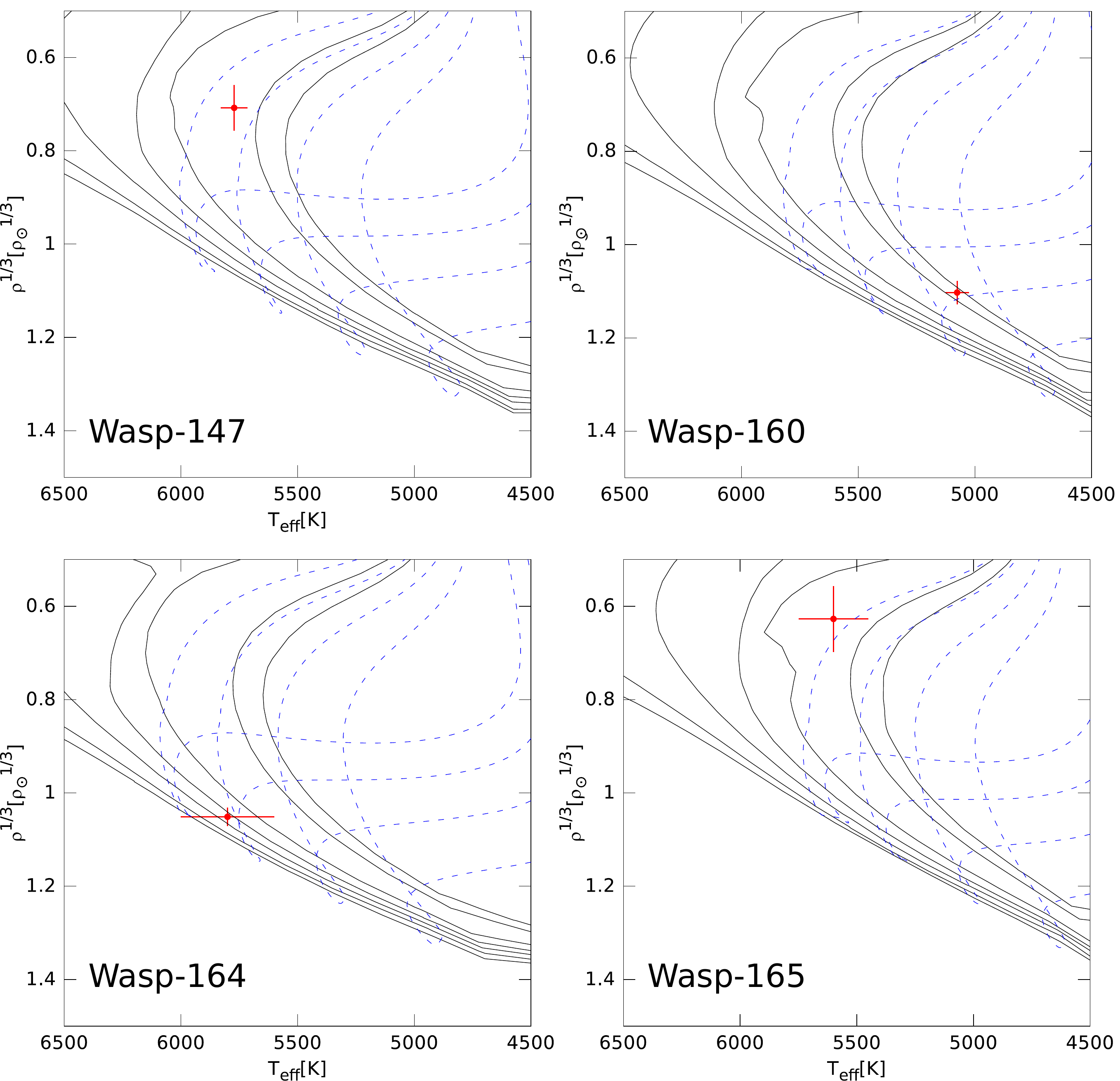}
    \caption{\label{fig:HRD}Placement of the planet host stars in the HR diagram. In each panel, theoretical models corresponding to the metallicity of that star 
    are shown. Solid black lines are representative of isochrones: from left to right 0.5, 1, 2, 3.2, 5, 10 and 12.6 Gyr models are represented. 
    Dashed blue lines are representative of evolutionary tracks: from left to right 1.1, 1, 0.9 and 0.8 $M_{\odot}$ models are represented.}
\end{figure*}

Uncertainties given by the code are simply internal, i.e. they are related to the interpolation scheme in use. Realistic uncertainties to be attributed to stellar parameters 
should take also theoretical model uncertainties into account. By comparing the results with two independent evolutionary models (namely PARSEC and CLES\footnote{%
Code Li\'{e}geois d'Evolution Stellaire}, \citealt{scuflaire08}), we find that systematics due to models can be estimated to be $\sim2\%$. In addition, helium content 
$Y$ influences theoretical models, but its quantity cannot be estimated from spectroscopy (at least in the case of solar-like stars). Given the uncertainty on $Y$, a 
further $\sim5\%$ should be added to the error budget. Fig.\ \ref{fig:HRD} shows the placement of the planet hosts in the HR diagram.

We also used the open source software {\sc
bagemass}\footnote{http://sourceforge.net/projects/bagemass} to calculate
the posterior mass distribution for each star  using the Bayesian method
described by \citet{Maxted15}. The models used in {\sc bagemass}
were calculated using the {\sc garstec} stellar evolution code
\citep{Weiss08} using as input the spectroscopically-derived {\teff} and $\mathrm{[Fe/H]}$ as well as the transit-derived {\rhoast} and orbital Period. 
The mass and age of the stars found are shown in
Table~\ref{bagemass_table}. They are in excellent agreement with the values
derived above for WASP-147, 164 and 165. For WASP-160B, {\sc bagemass} favors a slightly younger 
age and higher mass. This is due to the input selection of {\sc bagemass} that includes the spectroscopically-determined stellar effective temperature instead of the Gaia color index.

\begin{table}
\caption{Stellar mass and age estimates obtained with {\sc bagemass}. The mean and standard deviation of
the posterior distributions are given together with the best-fit values in
parentheses. \label{bagemass_table} }
\begin{tabular}{lrr}
\hline
Star & \multicolumn{1}{l}{Mass [$M_{\rm \odot}$]}& 
\multicolumn{1}{l}{Age [Gyr]}\\
\hline
\hline
WASP-147 &  1.08 $\pm$ 0.07 (1.05) & 8.4 $\pm$ 1.9 (9.1) \\
WASP-160B &  0.98 $\pm$ 0.04 (1.01) & 3.7 $\pm$ 2.3 (1.5) \\
WASP-164 &  0.96 $\pm$ 0.07 (1.02) & 4.7 $\pm$ 3.3 (2.0) \\
WASP-165 &  1.17 $\pm$ 0.09 (1.18) & 7.3 $\pm$ 2.2 (7.3) \\
\hline
\end{tabular}
\end{table} 

\subsection{The WASP-160 binary}
\label{sec:W160A}
Due to the low resolution of the WASP instrument, photometry of WASP-160B was blended with a second, slightly brighter, source. While exploratory RV observations and transit 
follow-up quickly identified the origin of the transit the to be the fainter star, we found that both objects possess near-identical systemic RVs, pointing towards them being 
physically associated. This is confirmed by consistent Gaia proper motion and parallax values for both objects. To derive the properties of WASP-160A, we retrieved the stellar 
properties from a spectral analysis and stellar evolution models as described in Section \ref{sec:specana} and \ref{sec:evol}. For the stellar evolution models, we used Gaia 
values for $G$, $G_{BP}-G_{RP}$ and $d$, and results from our spectroscopic analysis for [Fe/H], $\log{g}$, and $v\sin{i}$ as inputs. As we have not detected any transiting 
planet around WASP-160A, no transit-derived value for $\rho_{\star}$ was available. Our handful of RV measurements of WASP-160A are stable within $\sim$40~{\ms}. We summarize 
the properties of WASP-160A and the WASP-160A+B binary in Table \ref{tab:W160A}. Both stars appear to have similar masses and early K spectral types and their projected separation 
of $28.478948 \pm 2.5 \times 10^{-5}$~arcsec translates into a physical distance of $8060 \pm 101$~au. As WASP-160B, also WASP-160A has a super-Solar metallicity, 
([Fe/H] = $0.19\pm0.09$), consistent within uncertainties with the value found for WASP-160B.
Even though we would expect the two object to be coeval, we find a slightly older age for WASP-160A from evolutionary models. While this reinforces the older age estimate 
for WASP-160A, the discrepancy found in our analysis between objects A and B is likely an artifact of the very limited available data on WASP-160A.

\begin{table}
\caption{Properties of WASP-160A and the WASP-160A+B binary \label{tab:W160A} 
$^a$ assuming a macroturbulence of $2.43 {\kms}$ from \citet{2014MNRAS.444.3592D}.
$^b$ from stellar evolutionary models, $^c$ From Gaia DR2 and CORALIE, $^d$ computed from 
parallaxes corrected for the -80 $\mu$arcsec offset found by \citet{Stassun18}.
}
\begin{tabular}{p{3.4cm}p{1.9cm}p{1.9cm}}
\hline
\hline
\multicolumn{3}{l}{WASP-160A stellar properties} \\
\hline
RA (J2000)    & \multicolumn{2}{c}{05 50 44.7} \\
DEC (J2000)   & \multicolumn{2}{c}{-27 37 05.0} \\
UCAC4 B [mag] & \multicolumn{2}{c}{$13.452 \pm0.04 $} \\
UCAC4 V [mag] & \multicolumn{2}{c}{$12.677 \pm0.02 $} \\
2MASS J [mag] & \multicolumn{2}{c}{$11.300 \pm 0.026 $} \\
2MASS H [mag] & \multicolumn{2}{c}{$10.937 \pm 0.024 $} \\
2MASS K [mag] & \multicolumn{2}{c}{$10.831 \pm 0.019 $} \\
GAIA DR2 ID   & \multicolumn{2}{c}{2910755484609594368} \\
GAIA G [mag]  & \multicolumn{2}{c}{12.5} \\
GAIA $G_{\rm BP} [mag]$ & \multicolumn{2}{c}{12.9} \\
GAIA $G_{\rm RP} [mag]$ & \multicolumn{2}{c}{11.9} \\
Spectral Type & \multicolumn{2}{c}{K0V} \\
\teff [K] & \multicolumn{2}{c}{$5300\pm150$} \\
\logg [cgs] & \multicolumn{2}{c}{$4.5\pm0.2$} \\
$\mathrm{[Fe/H]}$ & \multicolumn{2}{c}{$0.19\pm0.09$} \\
\Vsini [\kms]$^a$& \multicolumn{2}{c}{$0.4\pm0.2$} \\
$\log{A({\rm Li})}$ &\multicolumn{2}{c}{ - } \\
Mass [{\Msolar}] & \multicolumn{2}{c}{$0.89\pm0.07$} \\
Radius [{\Rsolar} & \multicolumn{2}{c}{$0.95\pm0.07$}  \\
Age [Gyr]$^b$ & \multicolumn{2}{c}{$11.2\pm1.6$} \\
\hline
\hline
\multicolumn{3}{l}{Binary properties$^c$} \\
\hline
 & WASP-160A & WASP-160B \\
\hline
Proper motion RA [mas]  & $26.85\pm0.03$ & $26.97\pm0.03$ \\
Proper motion DEC [mas] & $-34.82\pm0.04$ & $-34.83\pm0.04$ \\
Parallax [mas] & $3.46\pm0.03$ & $3.44\pm0.02$ \\
Distance [pc]$^d$ & $282\pm5$ & $284\pm5$ \\
System RV [{\kms}] & $-6.1587\pm0.0075$& $-6.1421\pm0.0012 $ \\
\hline
\multicolumn{2}{l}{Position angle of B with respect to A [deg]} & -129.96 \\
Separation [arcsec]& \multicolumn{2}{r}{$28.478948 \pm 2.5 \times 10^{-5}$} \\
Separation [au]& \multicolumn{2}{r}{$8060 \pm 101$} \qquad \\
\hline
\end{tabular}
\end{table} 

\section{System parameters}
\label{sec:sys}

\subsection{Modeling approach}
\label{sec:mod}
We carried out a global analysis of all follow-up photometry and RVs for each planetary system using the Markov Chain Monte Carlo (MCMC) framework described
by \citet{Gillon12a}. In short, our model consists of a Keplerian for the RVs and the prescription of \citet{Mandel02} for transit light curves. Next to the 
fitted (``jump'') parameters listed in Table \ref{tab:par}, the code allows for the inclusion of parametric baseline models in the form of polynomials (up to 4th order) 
when fitting transit light curves. We tested a wide range of baseline models, including dependencies on time, airmass, stellar FWHM, coordinate shifts and sky background, 
and finally selected the appropriate solution for each light curve via Bayes factor comparison \citep[e.g.][]{Schwarz78}. The best baseline models are listed in Table \ref{tab:obs}. 
For all objects, we tested for a non-zero eccentricity by running two sets of global analyses: one while fixing the eccentricity to zero and one fitting for it by including 
$\sqrt{e}\sin{\omega}$ and $\sqrt{e}\cos{\omega}$ as jump parameters. We found no significant evidence for a non-circular orbit for any of our targets.

To estimate excess noise, we calculated the $\beta_{\mathit{r}}$ and $\beta_{\mathit{w}}$ \citep{Winn08,Gillon10a} factors 
that compare the rms of the binned and unbinned residuals and multiplied our error bars by their product before deriving the final parameter values.
We find no excess ``jitter'' noise in the RVs and thus do not adapt the RV errors. We adopted a quadratic stellar limb-darkening law using 
coefficients interpolated from the tables by \citet{Claret11}. To use the most appropriate input stellar parameters, we use the information extracted from the 
stellar spectra, paired with the Gaia DR2 \citep{gaia18} data as described in Section \ref{sec:stel}. After carrying out an initial analysis to measure transit-based
stellar mean densities needed to constrain the evolutionary models, we placed normal priors on {\Mast}, $\mathrm{[Fe/H]}_{\ast}$ and {\teff} centered on the values 
inferred in Section \ref{sec:stel}, with a width of the quoted $1-\sigma$ uncertainties. The four objects under study are revealed to be gas giants, two of them being 
classical hot Jupiters, while two have masses near that of Saturn. 

\begin{table*}
\centering                        
\caption{\label{tab:par}Planetary and stellar parameters from a global MCMC analysis. 
\newline $^a$Equilibrium temperature, assuming $A_B=0$ and $F=1$ \citep{Seager05}.}
\begin{tabular}{p{5.2cm}cccc}       
\hline\hline 
 Parameter & WASP-147 & WASP-160B & WASP-164 & WASP-165  \\
\hline
 \multicolumn{5}{l}{Jump parameters} \T  \\
\hline
 Transit depth, $ \Delta F$ \T                                       & $ 0.00640_{-0.00059}^{+0.00062} $      &  $ 0.01663_{-0.00038}^{+0.00043}   $    & $ 0.01542\pm0.00033 $                & $ 0.00544_{-0.00059}^{+0.00067}  $ \\
 $ b' = a*\cos(i_{p})$ $[R_{\ast}]$ \T                               & $ 0.31_{-0.21}^{+0.19} $               &  $ 0.20_{-0.12}^{+0.10}            $    & $ 0.8216_{-0.0091}^{+0.0084} $       & $ 0.53_{-0.22}^{+0.11}  $          \\
 Transit duration, $ T_{14}$ [d]      \T                             & $ 0.1831_{-0.0033}^{+0.0041} $         &  $ 0.11851_{-0.00078}^{+0.00094}   $    & $ 0.06682_{-0.00084}^{+0.00085} $    & $ 0.1740_{-0.0034}^{+0.0040} $     \\
 Mid-transit time, [BJD] - 2450000                     \T            & $ 6562.5950\pm0.0013 $                 &  $ 7383.65494\pm0.00021            $    & $ 7203.85378\pm0.00020 $             & $ 7649.71142\pm0.00093 $           \\
 Period, $ P$ [d]                                    \T              & $ 4.60273\pm0.000027 $                 &  $ 3.7684952 \pm0.0000035          $    & $ 1.7771255\pm0.0000028 $            & $ 3.465509\pm0.000023 $            \\
 $ K_2=K\sqrt{1-e^2}P^{1/3}$~[$\mathrm{ms^{-1}d^{1/3}}$]\T           & $ 54.4_{-4.6}^{+4.7} $                 &  $ 62.1_{-9.5}^{+9.2}              $    & $ 445\pm15 $                         & $ 115\pm16 $                       \\
 Stellar mass, $ M_{\ast} $ [{\Msolar}]              \T              & $ 1.044_{-0.073}^{+0.070} $            &  $ 0.87_{-0.068}^{+0.071}          $    & $ 0.946_{-0.071}^{+0.067} $          & $ 1.248_{-0.070}^{+0.072} $        \\
 Stellar eff. temperature,  {\teff}  [K]      \T                     & $ 5702\pm100 $                         &  $ 5298\pm99                       $    & $ 5806_{-200}^{+190} $               &  $ 5599\pm150 $                    \\
 Stellar metallicity, $ \mathrm{[Fe/H]}_{\ast} $     \T              & $ 0.092_{-0.071}^{+0.069} $            &  $ 0.27\pm0.10                     $    & $ -0.01_{-0.20}^{+0.19} $            &  $ 0.33\pm0.13 $                   \\
 $ c_{1,\rm V}=2u_{1,\rm V}+u_{2,\rm V} $            \T              & $ 1.198_{-0.050}^{+0.051} $            &                                          &                                      &                                    \\
 $ c_{2,\rm V}=u_{1,\rm V}-2u_{2,\rm V} $            \T              & $ -0.014\pm0.040 $                     &                                          &                                      &                                    \\
 $ c_{1,\rm R}=2u_{1,\rm R}+u_{2,\rm R} $            \T              &                                        &                                          & $  1.042_{-0.082}^{+0.081} $         &                                    \\
 $ c_{2,\rm R}=u_{1,\rm R}-2u_{2,\rm R} $            \T              &                                        &                                          & $ -0.178_{-0.056}^{+0.053} $         &                                    \\
 $ c_{1,\rm r'}=2u_{1,\rm r'}+u_{2,\rm r'} $         \T              & $  1.079_{-0.044}^{+0.046} $           &  $ 1.200\pm0.042 $                       & $  1.022_{-0.074}^{+0.077} $         &                                    \\
 $ c_{2,\rm r'}=u_{1,\rm r'}-2u_{2,\rm r'} $         \T              & $ -0.133_{-0.032}^{+0.033} $           &  $ 0.093\pm0.036 $                       & $ -0.178_{-0.059}^{+0.058}  $        &                                    \\
 $ c_{1,\rm NGTS}=2u_{1,\rm NGTS}+u_{2,\rm NGTS} $   \T              &                                        &                                          &                                      &  $ 1.09\pm0.10 $                   \\
 $ c_{2,\rm NGTS}=u_{1,\rm NGTS}-2u_{2,\rm NGTS} $   \T              &                                        &                                          &                                      &  $ 0.02\pm0.11 $                   \\                                                                                                                                                                                                     
 $ c_{1,\rm I+z'}=2u_{1,\rm I+z'}+u_{2,\rm I+z'} $   \T              &                                        &  $ 0.94\pm0.10  $                         & $  0.88\pm0.10 $                     &                                    \\
 $ c_{2,\rm I+z'}=u_{1,\rm I+z'}-2u_{2,\rm I+z'} $   \T              &                                        &  $ -0.07\pm0.11 $                        & $ -0.14_{-0.11}^{+0.10}  $           &                                    \\
 $ c_{1,\rm BB}=2u_{1,\rm BB}+u_{2,\rm BB} $         \T              & $ 0.99\pm0.10 $                        &                                          & $  0.99\pm0.10 $                     &                                    \\
 $ c_{2,\rm BB}=u_{1,\rm BB}-2u_{2,\rm BB} $         \T              & $ 0.00\pm0.11 $                        &                                          & $ -0.07\pm0.11 $                     &                                    \\
\hline                                       
 \multicolumn{5}{l}{Deduced parameters} \T \\
\hline                                       
 RV amplitude, $ K $ [{\ms}] \T                                      & $ 32.7\pm2.8 $                         &  $ 39.9_{-6.1}^{+5.9} $                  & $ 367\pm12 $                        &  $ 76\pm10 $                       \\
 RV zero point (pre-upgrade), $\gamma_{\rm COR1} $ [{\kms}]    \T    & $ -1.63849\pm0.00061 $                 &                                          &                                      &                                    \\
 RV zero point, $\gamma_{\rm COR2} $ [{\kms}]                  \T    & $ -1.617011_{-0.000066}^{+0.000069} $  &  $ -6.1421\pm0.0012 $                    & $ 12.262676_{0.000091}^{+0.000094}$  &  $ 25.6557\pm0.0012 $              \\
 Planetary radius, $ R_{p} $ [{\Rjup}]                         \T    & $ 1.115_{-0.093}^{+0.14} $             &  $ 1.090_{-0.041}^{+0.047} $             & $ 1.128_{-0.043}^{+0.041} $          &  $ 1.26_{-0.17}^{+0.19} $          \\
 Planetary mass, $ M_{p} $ [{\Mjup}]                           \T    & $ 0.275_{-0.027}^{+0.028} $            &  $ 0.278_{-0.045}^{+0.044} $             & $ 2.13_{-0.13}^{+0.12} $             &  $ 0.658_{-0.092}^{+0.097} $       \\
 Planetary mean density, $ \rho_{p} $ [{\rhojup}]              \T    & $ 0.198_{-0.060}^{+0.061} $            &  $ 0.214_{-0.038}^{+0.039} $             & $ 1.48_{-0.13}^{+0.15} $             &  $ 0.33_{-0.11}^{+0.19} $          \\
 Planetary grav. acceleration, $ \logg_{p} $ [cgs]             \T    & $ 2.74_{-0.11}^{+0.08} $               &  $ 2.763_{-0.077}^{+0.066} $             & $ 3.619_{-0.028}^{+0.029} $          &  $ 3.01_{-0.13}^{+0.14} $          \\
 Planetary eq. temperature, $ T_{eq} $ [K]$^a$                 \T    & $ 1404_{-43}^{+69} $                   &  $ 1119_{-23}^{+25} $                    & $ 1610_{-53}^{+58} $                  &  $ 1624_{-89}^{+93} $              \\
 Orbital semi-major axis, $ a $ [au]                           \T    & $ 0.0549_{-0.0013}^{+0.0012} $         &  $ 0.0452\pm0.0012$                      & $ 0.02818_{-0.00072}^{+0.00065} $    &  $ 0.04823_{-0.00092}^{+0.00091} $ \\
 $ a/R_{\ast} $                                                \T    & $ 8.29_{-0.74}^{+0.40} $               &  $ 11.25_{-0.31}^{+0.19} $               & $ 6.50\pm0.13$                       &  $ 5.93_{-0.55}^{+0.67} $          \\
 $ R_{p}/R_{\ast} $                                            \T    & $ 0.0800\pm0.0038 $                    &  $ 0.1290_{-0.0015}^{+0.0017}$           & $ 0.1242\pm0.0013$                   &  $ 0.0738_{-0.0041}^{+0.0044} $    \\ 
 Inclination, $ i_{p} $ [deg]                                  \T    & $ 87.9_{-1.6}^{+1.5} $                 &  $ 88.97_{-0.57}^{+0.63} $               & $ 82.73_{-0.21}^{+0.22} $            &  $ 84.9_{-1.7}^{+2.5}  $           \\
 Eccentricity, $ e $                                           \T    &  0 ($<0.19$ at $1\sigma$)              &  $ 0$ ($<0.22$ at $1\sigma$)             &  $0$ ($<0.09$ at $1\sigma$)          &  $0$ ($<0.14$ at $1\sigma$)        \\
 Stellar radius, $ R_{\ast} $ [{\Rsolar}]                      \T    & $ 1.429_{-0.076}^{+0.14} $             &  $ 0.868_{-0.028}^{+0.031} $             & $ 0.932_{-0.030}^{+0.028} $          &  $ 1.75\pm0.18 $                   \\
 Stellar mean density, $ \rho_{\ast} $ [{\rhosun}]             \T    & $ 0.361_{-0.088}^{+0.054} $            &  $ 1.34  _{-0.11}^{+0.07}  $             & $ 1.165_{-0.067}^{+0.074} $          &  $ 0.233_{-0.059}^{+0.089} $       \\ 
 Limb-darkening coefficient, $ u_{1,\rm V} $                   \T    & $ 0.477\pm0.027 $                      &                                          &                                      &                                    \\
 Limb-darkening coefficient, $ u_{2,\rm V} $                   \T    & $ 0.245\pm0.018 $                      &                                          &                                      &                                    \\
 Limb-darkening coefficient, $ u_{1,\rm R} $                   \T    &                                        &                                          &  $ 0.381\pm0.040 $                   &                                    \\
 Limb-darkening coefficient, $ u_{2,\rm R} $                   \T    &                                        &                                          &  $ 0.280_{-0.020}^{+0.021} $         &                                    \\
 Limb-darkening coefficient, $ u_{1,\rm r'} $                  \T    & $ 0.405_{-0.023}^{+0.024} $            &  $ 0.499  \pm0.022 $                     &  $ 0.373\pm0.038 $                   &                                    \\
 Limb-darkening coefficient, $ u_{2,\rm r'} $                  \T    & $ 0.269\pm0.015 $                      &  $ 0.203  \pm0.017 $                     &  $ 0.276_{-0.023}^{+0.022} $         &                                    \\
 Limb-darkening coefficient, $ u_{1,\rm NGTS} $                \T    &                                        &                                          &                                      &  $ 0.440\pm0.056 $                 \\ 
 Limb-darkening coefficient, $ u_{2,\rm NGTS} $                \T    &                                        &                                          &                                      &  $ 0.210\pm0.058 $                 \\
 Limb-darkening coefficient, $ u_{1,\rm I+z'} $                \T    &                                        &  $ 0.363  \pm0.053 $                     &  $ 0.323_{-0.047}^{+0.048} $         &                                    \\
 Limb-darkening coefficient, $ u_{2,\rm I+z'} $                \T    &                                        &  $ 0.215  \pm0.056 $                     &  $ 0.232_{-0.047}^{+0.049} $         &                                    \\
 Limb-darkening coefficient, $ u_{1,\rm BB} $                  \T    & $ 0.398\pm0.053 $                      &                                          &  $ 0.384_{-0.048}^{+0.047} $         &                                    \\
 Limb-darkening coefficient, $ u_{2,\rm BB} $                  \T    & $ 0.196\pm0.055 $                      &                                          &  $ 0.224_{-0.048}^{+0.051} $         &                                    \\
\hline
\end{tabular}
\end{table*}

\subsection{WASP-147}
WASP-147b is a Saturn-mass ($M = 0.27 \Mjup$) planet orbiting a G4 star with a period of 4.6 days. The system appears to be old, with the 1.04 \Msolar \ host having started to evolve off the main 
sequence. Stellar evolutionary modeling places the star's age at $8.5 \pm 0.8$~Gyr, and its old age is corroborated by a low Lithium abundance that is in accordance with measurements for stars aged 
2~Gyrs or more \citep{Sestito05} and the absence of activity indicators such as excess RV stellar noise and rotational variability. The planet is one of the more strongly irradiated 
planets of its mass range. Considering the mass--incident flux plane shown in Fig.\ \ref{fig:MRP}, WASP-147b is located near the inner tip of the triangular sub-Saturn desert 
\citep{Mazeh05,Szabo11,Mazeh16}, which appears to be created by erosion of planetary atmospheres due to stellar irradiation \citep{Lammer03,Baraffe06}. Being a low-mass, low-density planet, 
WASP-147b is a good target for transmission spectroscopy. One atmospheric scale height translates to a predicted change in the transit depth of 249~ppm, well within the precision of 
ground- and space-based transmission spectra \citep[e.g.][]{Kreidberg15,Sing16,Lendl17b,Sedaghati17}. 

\subsection{WASP-160B}

Similar to WASP-147b, WASP-160Bb is also a near Saturn-mass ($M_{p} = 0.28 \Mjup$) object, however this planet orbits a cooler K0V star in a wide (28.5~arcsec) near equal-mass binary 
with a period of 3.8 days. The stellar age is estimated to be approximately 10~Gyrs from evolutionary models, and a non-detection of Lithium supports the object's old age. 
In contrast to WASP-147, the later-type WASP-160B still resides firmly on the main sequence. Both planets share near-identical mass and radius, however WASP-160Bb orbits a very
metal-rich star ($\mathbf{\mathrm{[Fe/H]=0.27\pm0.1}}$), while WASP-147's metallicity is near-Solar. WASP-160Bb receives less stellar irradiation than 
the bulk of hot Jupiters (see Fig.\ \ref{fig:MRP}), which translates to a moderate equilibrium temperature of approx.\ 1100 K. Prospects for studying this object's atmospheric transmission 
spectrum are excellent, as one atmospheric scale height translates to a radius variation of 338~ppm. To date, the only hot Jupiter orbiting a more metal-rich host with a characterized 
transmission spectrum is XO-2 \citep{Burke07,Sing11b}, for which both Na and K have been detected. 

\subsection{WASP-164}

WASP-164b is a massive ($M_p = 2.13 \Mjup$) planet on a close-in orbit around a G2V star. Owing to the limited S/N of the follow-up spectra obtained, the stellar properties are somewhat more uncertain than 
for the other objects presented here. However, rotational modulation with a period of $17.81 \pm 0.03$~days seen in the survey light curve indicates a relatively young age of $2.3_{-0.5}^{+1}$~Gyr and 
evolutionary models confirm the star being located on the main sequence. The planet is rather massive compared to the bulk of hot Jupiters and orbits its host star with a short orbital period
of 1.78~days at a separation of 2.74 times its Roche limit. This is consistent with expectations from high-eccentricity migration mechanisms \citep{Ford06}. Being rather massive and close-in, the planet may 
be further undergoing orbital decay through tidal dissipation \citep[e.g.][]{Rasio96b,Levrard09, Matsumura10}. Using a value of $\log{Q'_{s}}=8.26\pm0.14 $ \citep{CollierCameron18}, 
we find a remaining lifetime of $13.0\pm4.6$~Gyr. The planetary orbit is thus not expected
to shrink drastically due to tidal dissipation over the star's main-sequence lifetime. The planet's close-in orbital period is however advantageous for the object's atmospheric characterization. Indeed, WASP-164b makes 
for a promising target for characterization via emission spectroscopy, with a predicted occultation depth in H band between 450~ppm (full redistribution) and 1125~ppm (immediate reradiation).

\subsection{WASP-165}
\begin{figure}
	\includegraphics[width=\columnwidth]{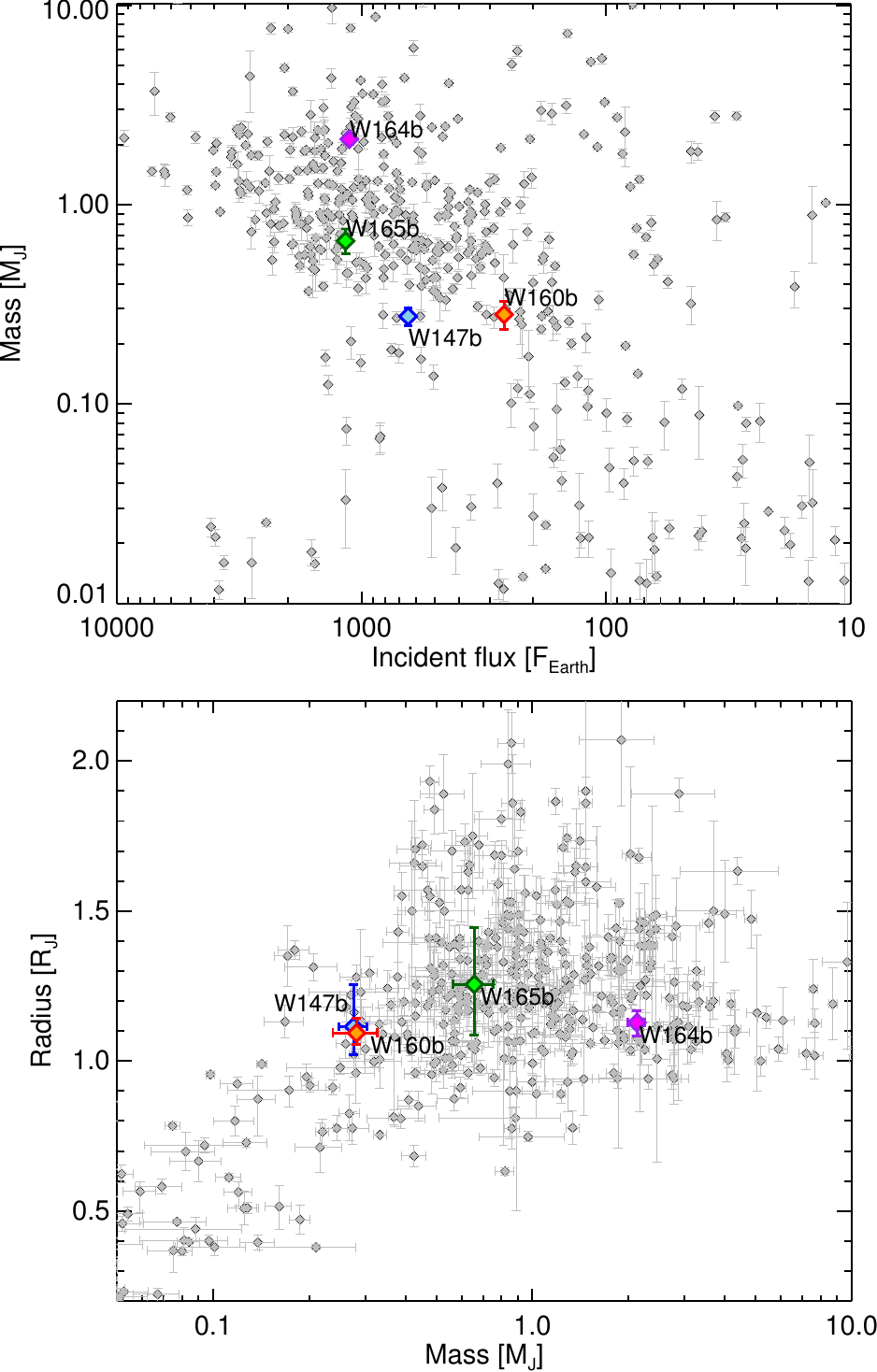}
    \caption{\label{fig:MRP}Top: Planetary masses against incident flux for known exoplanets. Only planets with well-measured masses and radii 
    (relative uncertainties smaller than 50\%) are shown. Our newly-discovered objects are shown in color and labelled. Bottom: Planetary mass-radius diagram. 
    Sample selection and designation of our targets as above.}
\end{figure}

WASP-165b is a classical hot Jupiter. With an orbital period of 3.4~days, it lies near the peak of the hot Jupiter period distribution while its mass ($M_P = 0.66 \Mjup$) and 
radius ($R_P = 1.26 \Rjup$) are typical for this class of planet (see Fig.\ \ref{fig:MRP}). With respect to planetary mass, radius and period, WASP-165b is near-identical to the extremely well-studied object HD~209458b \citep{Mazeh00,Charbonneau00,Henry00}, however WASP-165b is more strongly irradiated as its $M_\ast=1.25 \Msolar$ G6 host has started evolving off the main sequence. The system's age estimate from evolutionary 
models is $4.8\pm1$~Gyr, albeit the object's Lithium abundance suggests a slightly younger age of 
$0.5 - 2$~Gyr. This discrepancy is most likely due to the limited S/N of our spectra. We consider the age estimate from evolutionary models as more accurate, as these contain the stellar mean density inferred from the high-precision follow-up light curves. As WASP-160B, also WASP-165 has a remarkably high metallicity of $\mathrm{[Fe/H]=0.33\pm0.13}$. 

\section{Summary}
\label{sec:sum}
We present the discovery of four transiting hot Jupiters by WASP-South. WASP-147b and WASP-160Bb are Saturn-mass planets with near-identical radii but contrasting 
stellar metallicities and planetary equilibrium temperatures: the 1400~K WASP-147b orbits a Solar-metallicity G4 star, while the 1100~K WASP-160Bb orbits a metal-rich K0 
star in a near equal-mass binary. Both objects are promising targets for atmospheric characterization via transmission spectroscopy. WASP-164b is a massive ($2.1\, \Mjup$) 
planet on a short (1.8~d) orbital period, and a potential target for emission spectroscopy. Finally, WASP-165b is a classical hot Jupiter orbiting a metal-rich host.

\section*{Acknowledgements}
WASP-South is hosted by the South African Astronomical Observatory 
and we are grateful for their ongoing support and assistance. 
Funding for WASP comes from consortium universities
and from the UK's Science and Technology Facilities Council.
TRAPPIST is funded by the Belgian Fund for Scientific  
Research (Fond National de la Recherche Scientifique, FNRS) under the  
grant FRFC 2.5.594.09.F, with the participation of the Swiss National  
Science Fundation (SNF). 
MG is a F.R.S.-FNRS Senior Research Associate. The research leading to these results has received 
funding from the European Research Council under the FP/2007-2013 ERC Grant Agreement 336480, 
from the ARC grant for Concerted Research Actions financed by the Wallonia-Brussels Federation, and from the Balzan Foundation.
L.D. acknowledges support from the Gruber Foundation Fellowship.
M.L. acknowledges support from the Austrian Research Promotion Agency (FFG) under project 859724 "GRAPPA". 




\bibliographystyle{mnras}
\bibliography{bbl.bib} 




\bsp	
\label{lastpage}
\end{document}